\documentclass[12pt,a4papper]{article}

\begin{document}

\title{Interplay of topology and quantization: topological energy quantization  in a cavity}
\author{Antonio F. Ra\~nada\thanks{E-mail: afr@fis.ucm.es}\\{\small Departamento de F\'{\i}sica Aplicada III, Universidad Complutense, 28040 Madrid, Spain}}
\date{1 November 2002}
\maketitle

\begin{abstract}
The interplay between quantization and topology is investigated in the frame of
 a topological model of electromagnetism proposed by the author. In that model,
the energy of electromagnetic radiation in a cubic cavity is
${\cal E}=(d/4)\hbar \omega $ where $d$ is a topological integer
index equal to the degree of a map between two orbifolds.
\end{abstract}

\centerline{To be published in Physics Letters A}

\bigskip
{\small


 \hspace{1.8cm} Energy quantization in a cavity; Planck law; Knot}

\newpage

\setcounter{page}{1}

\section{Introduction: Adiabatic and topological invariants}
The idea of adiabatic invariant was much used in the old quantum theory to understand the quantization of the radiation in a cavity, specially by Einstein and Ehrenfest. At the Solvay Conference in 1911, Einstein answered a question raised by Lorentz with the statement ``If the length of a pendulum is changed infinitely slowly, its energy remains equal to $h\nu$ if it was originally $h\nu$." More or less at the same time, Ehrenfest was puzzled by a paradox: ``Wien displacement law is wholly derived from classical foundations [but is] unshaken in the midst of ... phenomena whose anticlassical character stood out ever more inexorably," in his own words. He also recognized that Wien law establishes a relation between two adiabatic invariants ${\cal E}_\nu /\nu$ and $\nu /T$, where ${\cal E}_\nu$ is the energy of the proper vibration with frequency $\nu$ and $T$ is the temperature. Following a suggestion by Einstein, Ehrenfest coined then the expression ``adiabatic principle"\cite{Jam66} to name the statement that ``in the course of an adiabatic [very slow] transformation, an allowed (or stationary) undeformed motion changes into an allowed deformed motion, while the adiabatic invariant retains its initial value". This principle was important because it made possible to determine the allowed motions of any periodic system of one degree of freedom if it can be transformed by adiabatic change in the sinusoidal oscillator. The idea was later extended to systems with more than one degree of freedom. Note that the number of photons in a cavity filled with monochromatic radiation  is ${\cal E}_\nu /\nu$ (times the Planck constant), so that it is the same kind of quantity that prompted Einstein above mentioned statement in 1911.

The modern version of the quantum adiabatic principle asserts that
a system that is initially in a stationary state labelled by a set
of quantum numbers will remain in a stationary state labelled by
the same quantum numbers if its environment changes adiabatically.
However, this invariance is compatible with a subtle and most
important variation discovered by M. Berry in 1984: the phase of
its wave function may change in an amount, the now very famous and
quoted Berry's phase, that had been overlooked until then
\cite{Ber84}. It is interesting and significant that its study
requires both adiabatic and topological considerations.

We are interested here in topological invariants, which are
different from adiabatic invariants although the two types share
some common traits. An adiabatic invariant remains constant under
slow changes of some parameters that characterize a system. A
topological invariant keeps the same value under imaginary smooth
deformations of the motion or of some parameters which do not
involve time. They are similar, both implying that something is
invariant when something changes, but the two kinds of change are
different: in the adiabatic case, they are time evolutions in the
limit of infinitely slow deformations, while in the topological
case they are characterized by the variation of parameters without
temporal meaning. A further difference is that topological
invariants are usually discrete numbers, as are some of the
numbers that characterize the state of quantum systems, the ones
to which Einstein and Ehrenfest applied the adiabatic principle at
the beginning of 20th century. This intriguing discretization of
the physical quantities inspired a deep and thought provoking
remark by Atiyah \cite{Ati90}: ``Both topology and quantum physics
go from the continuous to the discrete." All this suggests that
the relative success of the adiabatic  principle in old quantum
theory was perhaps due to the common elements of adiabatic and
topological invariants.

The quantization of the energy $\cal E$ of the monochromatic
radiation in a cubic cavity is studied here in the frame of a
topological model of electromagnetism  (TME from now on) proposed
by the author.  As is shown here, the TME predicts that ${\cal
E}=(d/4)\hbar \omega $, where $d$ is a topological integer index
equal the degree of a map between a spacetime orbifold and a field
orbifold. The TME is summarized in section 2 \cite{Ran90a,Ran92}.
In section 3 and 4, the properties of a normal mode of the
electromagnetic field in a cubic cavity are reviewed, with
emphasis on its symmetries, which allow it to be defined in a
spacetime orbifold. Section 5 states the main result, the
principal conclusion being that the TME  offers a new and
promising approach to study the relation between topology and
quantization in the case of the electromagnetic field.

\section{The topological model of electromagnetism}
This section summarizes the basic elements of the TME proposed by the author,  which is
locally equivalent to Maxwell's standard theory but implying furthermore some
topological quantization conditions with interesting physical
meaning \cite{Ran89}-\cite{Ran98a} (\cite{Ran01} is a review where all
the basic details are explained but note that, the results of
this paper being new, they are not included there). The TME makes use of two fundamental complex scalar fields
$(\phi ,\theta )$ the level curves of which coincide with the
magnetic and electric lines, respectively, each one of these lines being labelled by the constant
value of the corresponding scalar. It turns out that
the set of magnetic and electric lines has very curious and
interesting topological properties.

The two scalars are assumed to have only one value at infinity,
which is equivalent to compactify the three-space to the sphere $S^3$. This
implies that they can be interpreted (via
stereographic projection) as maps $S^3\rightarrow S^2$, which can be classified in
homotopy classes and, as such, be characterized by the value of
the Hopf index $n$. It can be shown that the two scalars have the
same Hopf index and that the magnetic (resp. electric) lines are
generically linked with the same linking number in the sense of
Gauss $\ell$. If $\mu$ is the multiplicity of the level curves
(i.e. the number of different magnetic (resp. electric) lines that
have the same label $\phi$ (resp. $\theta$)), then $n=\ell \mu
^2$; the Hopf index can thus be interpreted as a generalized
linking number if we define a line as a level curve with $\mu$
disjoint components.

An important feature of the model is that the Faraday 2-form ${\cal F}=\frac{1}{2} F_{\mu \nu}dx^\mu \wedge
dx^\nu$ and its dual $*{\cal F}=\frac{1}{2}\, ^*\!F_{\mu \nu}dx^\mu \wedge
dx^\nu$ are the two pull-backs of $\sigma$, the area 2-form in $S^2$, by $\phi$ and $\theta$, i. e.
\begin{equation}
{\cal F}=-\sqrt{a}\,\phi ^*\sigma ,\;\;\;*{\cal F}=\sqrt{a}\,\theta ^*\sigma ,
\label{10}
\end{equation}
where $a= \sqrt{\hbar c\epsilon _0}$, in SI units ($\hbar
,c,\epsilon _0$ being the Planck constant, the light velocity and
the vacuum permittivity).  Natural units will be used here, so
that $a=1$. As a consequence the two maps are dual to one another
in the sense that
\begin{equation}
*(\phi ^*\sigma )=-\theta ^*\sigma ,
\label{20}
\end{equation}
* being the Hodge or duality operator. Curiously enough, the existence
of two maps satisfying (\ref{20}) guarantees that both $\cal F$
and $*\cal F$ obey the Maxwell equations in empty space without
the need of any other requirement. We will note ${\cal F}\equiv
({\bf E}, {\bf B})$, $^*\!{\cal F}\equiv (-{\bf B}, {\bf E})$.

The electromagnetic fields obtained as in (\ref{10}) are
called ``electromagnetic knots". They are radiation fields , i.e. they verify the condition ${\bf E}\cdot {\bf B}=0$. Note that, because of the Darboux theorem, any electromagnetic field in empty space can be expressed locally as the sum of two radiation fields.

As stated before, the TME is locally equivalent to Maxwell standard theory \cite{Ran01,Ran92,Ran95b}. However, its difference from the global point of view has interesting consequences, as are the following two topological quantizations:

i) In the TME, the electric charge of any point particle must necessarily be equal to an integer multiple of the fundamental value $q_0=\sqrt{\hbar c}$, i.e.  $q_0=1$ in natural units. Furthermore, if a charge has $m$ fundamental units $q=mq_0$, then $m$ is equal to the degree of the map $\theta ^\prime : \Sigma \rightarrow S^2$, the restriction of $\theta$ to $\Sigma$, this one being any closed surface enclosing the charge. Consequently, there are exactly $m$ lines converging to or diverging from the charge having any prescribed value of $\theta$ as their label. Note that $q_0=3.3 \,e=5.29 \times 10^{-19}$ C \cite{Ran95b,Ran98a}. In a previous paper it was suggested that $q_0$ might be the infinite energy limit of the electron charge $e$ \cite{Ran99d}, i.e. the bare charge. An intriguing feature of the model is that the hypothetical magnetic charges are also quantized with the same fundamental charge $q_0$.

ii) The electromagnetic helicity $\cal H$ is also quantized
\begin{equation}
{\cal H}= \frac{1}{2}\, \int _{R^3} \left({\bf A}\cdot {\bf B}+{\bf C}\cdot {\bf E}\right) \,d^3r =n,
\label{40}
\end{equation}
where ${\bf B}=\mbox{\boldmath$\nabla$} \times {\bf A}$, ${\bf E}= \mbox{\boldmath$\nabla$} \times {\bf C}$,
 the integer $n$ being equal to the common value of the Hopf indices of $\phi$ and $\theta$ (this is in natural units, in physical units, the right-hand side of (\ref{40}) would be $n\hbar c$). Note that  ${\cal H}=N_R -N_L$, where $N_R$ and $N_L$ are the classical expressions of the number of right- and left-handed photons contained in the field (i.e. ${\cal H}=N_R-N_L=\int d^3k (\bar{a}_Ra_R-\bar{a}_La_L)$,  $a_R({\bf k}),a_L({\bf k})$ being Fourier transforms of $A_\mu$ in the classical theory, but creation and annihilation operator in the quantum version). This implies that
\begin{equation}
n= N_R-N_L,
\label{50}
\end{equation}
which is a curious relation between the Hopf index (i.e. the
generalized linking number) of the classical field and the
classical limit of the difference $N_R-N_L$. This difference has a
clear topological meaning, what is
attractive from the intuitive physical point of view. The present work
shows that the rich topological structure of the model embodies a
third quantization also: the quantization of the energy of
monochromatic radiation in a cubic cavity.

To complete this section, a final remark is necessary.
The Faraday 2-form and its dual generated by the pair $(\phi ,\, \theta )$ can be written as
\begin{eqnarray}
{\cal F}&=&ds\wedge dp, \;\;\;\;\mbox{with }\; p=1/(1+|\phi |^2),\; s=\arg (\phi )/2\pi
\label{30a}\\
^*{\cal F}&=&dv\wedge du,\;\;\;\;\mbox{with }\;  v=1/(1+|\theta |^2),\, u=\arg (\theta )/2\pi ,
\label{30b}
\end{eqnarray}
so that $\phi =\sqrt{(1-p)/p}\, e^{i2\pi s}$ and $\theta =\sqrt{(1-v)/v}\, e^{i2\pi u}$.

This implies that the magnetic and electric fields have the form
\begin{eqnarray}
{\bf B} &=&\mbox{\boldmath$\nabla$} p\times \mbox{\boldmath$\nabla$} s
=\left( \partial _0u\mbox{\boldmath$\nabla$} v-\partial
_0v\mbox{\boldmath$\nabla$} u\right)   \nonumber \\
{\bf E} &=&  \mbox{\boldmath$\nabla$} u\times \mbox{\boldmath$\nabla$} v
=\left(\partial _0s\mbox{\boldmath$\nabla$} p-\partial _0p\mbox{\boldmath$\nabla$}
s\right)
\label{3.17}
\end{eqnarray}
The quantities $(p,s)$ and $(v,u)$ are called {\em Clebsch variables} of the fields ${\bf B}$ and ${\bf E}$, respectively (or of the scalars $\phi$ and $\theta$ as well). Note that $\phi$ and $\theta$ are not uniquely determined by the magnetic and electric fields. Indeed, a different pair  defines the same fields ${\bf E},\, {\bf B}$ if the corresponding Clebsch variables $(P,S)$, $(V,U)$   can be obtained  through a canonical transformation $(p,s) \rightarrow (P,S)$ or $(v,u) \rightarrow (V,U)$. However, the canonical transformation must satisfy two conditions: (i) $0\leq P,V\leq 1$ and (ii) $S,\, U$ must be arguments of complex numbers in units of $2\pi$, {\em i. e.}, they can be multivalued but their change along a closed curve must be an integer. Changes of Clebsch variables will be made later.

\section{Electromagnetic radiation in a cavity}
Let us consider a cubic cavity $\cal C$ with side $\pi$ ($0\leq x,y,z \leq \pi$), in which there is electromagnetic radiation in equilibrium with the walls.
To obtain the expression of the field inside the cavity, one has to solve the Maxwell equations with the boundary conditions ${\bf E}\times {\bf n} =0$, ${\bf B}\cdot {\bf n} =0$,
{\bf n} being a vector normal to the wall of the cavity ${\cal
S}=\partial {\cal C}$.

We are interested here in radiation or singular solutions (i.e. with ${\bf E}\cdot {\bf B}=0$), the ones that can correspond to electromagnetic knots. By choosing suitably the gauge, we can take $A^0=0$, after which the corresponding normal modes characterized by the triplet of integers $k_1,k_2,k_3$ can be expressed as
\begin{eqnarray}
A^0&=& 0,\;\;\;\;
A^1= Ae_{1x}\cos \omega t \cos k_1x \sin k_2y \sin k_3z, \label{70} \\
A^2&=& Ae_{1y}\cos \omega t\sin k_1x \cos k_2y \sin k_3z,
A^3=  Ae_{1z}\cos \omega t\sin k_1x \sin k_2y \cos k_3z \nonumber
\end{eqnarray}
where $\omega = |{\bf k} |=\sqrt{k_1^2+k_2^2+k_3^2}$, and $({\bf e}_1,{\bf e}_2,{\bf k}/k)$ being three orthonormal vectors. The electric and magnetic fields are ${\bf E}=-\partial _t{\bf A}$, ${\bf B}=\nabla \times {\bf A}$, their well known expressions being
\begin{eqnarray}
E_i &=&\omega Ae_{1i}\sin \omega t\cos k_ix_i\sin k_jx_j\sin k_kx_k,\nonumber \\
B_i&=&\omega Ae_{2i}\cos \omega t\sin k_ix_i\cos k_jx_j\cos k_kx_k,
\label{80}
\end{eqnarray}
where $(ijk)$ is a permutation of $(123)$, no summation being
implied over repeated indices.

{\bf Symmetry properties of the electric and magnetic fields of
the normal modes.}  First note that the electric and magnetic
fields corresponding to (\ref{70}) are periodic in any of the
coordinates $x_i$ and in time $t$, so that ${\bf E}(x_i+ 2\pi )=
{\bf E}(x_i),\, {\bf B}(x_i+ 2\pi )={\bf B}(x_i) $, ${\bf E} (t+
\tau ) ={\bf E}(t),\; {\bf B} (t+ \tau ) ={\bf B}(t)$. Moreover
they verify the following symmetry relations as is easy to show
\begin{eqnarray}
E_i (-x_i)=+E_i(x_i), &\;\;& E_i(-x_j)=-E_i(x_j),\nonumber \\
 B_i (-x_i)=-B_i(x_i), &\;\;& B_i(-x_j)=+B_i(x_j),\label{100}\\
{\bf E}(-t )=-{\bf E}(t ), &\;\;& {\bf B}(-t)={\bf B}(t).\label{110}
\end{eqnarray}
with $i\neq j$ (the space or time coordinates that do not appear are not changed.)
Although (\ref{110})
refers to the particular solution that we are considering, it is
easy to see that this election implies no loss of generality, since it is always verified after a convenient change of the
origin of the time variable.
Note that the symmetries (\ref{100})-(\ref{110}) suggest that the electromagnetic field can be considered to be defined in an orbifold. This will be important later.

The energy in the cavity is equal to
\begin{equation}
{\cal E} =\frac{1}{2}\int _{\cal C} \left(E^2 +B^2\right)\,d^3r= \omega ^2 A^2\frac{\pi ^3}{16},
\label{120}
\end{equation}
 and take continuous values depending on $A$, according to classical physics. On the other hand, quantum physics is based on the Planck-Einstein relation, which in this case is written (with $\hbar =1$)
\begin{equation}
{\cal E} = n\omega .
\label{140}
\end{equation}

\section{The topological model in the cavity}

The electric and magnetic fields of the mode $(k_1,k_2,k_3)$ given by (\ref{70}) are periodic in $x,y,z$ with periods $2\pi /k_i$ in each coordinate, and in time $t$ with period  $\tau =2\pi /\omega$.  Consequently, they are defined in the cartesian product of a 3-torus $T^3$ (the cube $0\leq x,y,z \leq 2\pi$ in which the opposite faces are identified), and the 1-torus $T^1$ with coordinate $t$ so that $0\leq t\leq \tau \equiv 0$. Let us note $T^4=T^3\times T^1$. We can then consider the pair $\phi ,\theta$ as a map
\begin{equation}
\eta \equiv \phi \times \theta :T^4\rightarrow S^2\times S^2,
\label{150}
\end{equation}
its degree being an integer $n$
\begin{equation}
\int _{T^4}\eta ^*\sigma =\int _{T^4}\phi ^*\sigma \wedge\theta
^*\sigma =n. \label{160}
\end{equation}
This can be written as
\begin{equation}
\int _{T^4} {\cal F}\wedge *{\cal F} =\int _{T^4} {E^2-B^2\over
2}\,d^4x  =n, \label{170}
\end{equation}
that gives a topological condition on the electromagnetic field.
However, this is not useful since the Maxwell equations imply that
$n=0$ always. Other topological numbers are the fluxes across the
faces of the cavity and other similar numbers, but they are again
zero. However, although the electromagnetic field is defined in
the torus $T^4$,  it has furthermore some additional symmetries
with important consequences.

{\bf Two new maps $\phi ^\prime$ and $\theta ^\prime $ and their
symmetry properties.} Since, as noted before, the normal modes can
be considered to be defined in a four dimensional torus $T^4$, we
will take scalar fields satisfying the symmetries $\phi (x_i)
=\phi (x_i+ 2\pi ),\;\; \theta (x_i) =\theta (x_i+ 2\pi ),\;\;\phi
(t+ \tau )=\phi (t),\;\;  \theta (t+ \tau )=\theta (t)$.
As noted at the end of section 2, the correspondence between the fields ${\bf B},\, {\bf E}$ and their Clebsch variables $(p,s)$, $(v,u)$ is not unique, there are different pairs of scalars that give the same electromagnetic field.

Using that freedom, we will define now two new maps $\phi ^\prime $ and $\theta ^\prime$, closely related to $\phi$ and $\theta$, which will be used in section 5 to prove the main result of this paper. They will be defined so that they satisfy the following symmetry relations (with $\tau =2\pi /\omega$)
\begin{equation}
\phi ^\prime (x_i) =\phi ^\prime (x_i+ 2\pi ),\;\; \theta ^\prime (x_i) =\theta ^\prime (x_i+ 2\pi ),\;\;\phi ^\prime (t+ \tau )=\phi ^\prime (t),\;\;  \theta ^\prime (t+ \tau )=\theta ^\prime (t),
\label{180}
\end{equation}
\begin{equation}
\phi ^\prime (-x_i) =\left(\bar{\phi}^\prime (x_i)\right)^{-1},\;\;\;\;  \theta ^\prime (-x_i) =\theta ^\prime (x_i),\
\label{190}
\end{equation}
\begin{equation}
\phi ^\prime ({\bf r}, \tau /2-t)=\left(\bar{\phi}^\prime ({\bf r}, t\right)^{-1},\;\;\;\;
\theta ^\prime ({\bf r},\tau /2- t)=\theta ^\prime ({\bf r}, t),
\label{195}
\end{equation}
\begin{equation}
 \phi ^\prime ({\bf r} ,t+\tau / 2)= \left(\bar{\phi}^\prime  ({\bf r} ,t)\right)^{-1},\;\;\;\; \theta ^\prime ({\bf r} ,t+\tau /2)= \left(\bar{\theta}^\prime  ({\bf r} ,t)\right)^{-1},
\label{200}
\end{equation}
\begin{equation}
\phi ^\prime({\bf r}  ,-t) =\phi ^\prime({\bf r}  ,t),\;\;\;\; \theta ^\prime ({\bf r} ,-t) = \left(\bar{\theta}^\prime  ({\bf r} ,t)\right)^{-1},
\label{210}
\end{equation}
Furthermore, the phases (or arguments) of the scalars are time independent.
 Note that (i) in (\ref{180}) and (\ref{190}), only one space coordinate is changed each time;
(ii) because of the space periodicity in (\ref{180}), equation
(\ref{190}) is equivalent to $\phi ^\prime (\pi -x_i)
=1/\bar{\phi}^\prime (\pi +x_i), \;\; \theta ^\prime (\pi -x_i)
=\theta ^\prime (\pi +x_i)$; and
 (iii) any of the three pairs of equations (\ref{195})-(\ref{210}) is a consequence of the other two.

Taken together, these equations mean that (a) both scalars are
periodic in the four spacetime coordinates, with period $2\pi$ in
$x_i$ and $\tau$ in time; (b) $\phi ^\prime$ changes into the
inverse of its complex conjugate $(\bar{\phi}^\prime )^{-1}$ under
reflections with respect to any plane $x_i=m\pi$ ($m$ being an
integer), while $\theta ^\prime $ remains invariant; (c) $\phi
^\prime$ changes into the inverse of its complex conjugate while
$\theta ^\prime$ remains invariant under reflection with respect
to the time $t=\tau /4$; (d) both scalars change to the inverse of
their complex conjugates after half a period; (e) $\phi ^\prime $
is even under time reversal while $\theta ^\prime$ changes into
the inverse of its complex conjugate.

It will be shown now that, given an expression for $\phi$, we can
find another $\phi ^\prime$ that generates the same
electromagnetic field and verifies the first equation (\ref{190})
by applying to $\phi$ two successive  transformations $T_1:\phi
\equiv \phi ^{\rm old}\rightarrow  \zeta$ and $T_2:\zeta
\rightarrow \phi ^{\rm new} $. This will be done as follows.

Because of the boundary condition ${\bf B}\times {\bf n}=0$, the magnetic lines contained in the walls
form a one-dimensional set and define a function $s=f(p)$
(possibly with several branches.)
 A simple way of changing the scalar $\phi$ without changing the
 electromagnetic field is to add a function $g(p)$ to $s$ in (\ref{30a}),
 so that $\phi ^{\rm old}$ changes to $\zeta =|\phi| e^{i2\pi S}$ with $S=s+g(p)$.
  It is clear that the Faraday  2-form in (\ref{30a}) and the vector fields in
   (\ref{3.17}) remain invariant. The transformation $T_1$ is defined to be of
  this type with $g(p)=-f(p)$. The new scalar $\zeta$ is obviously real in the walls
   of the cavity (but generically not outside or inside). As $\mbox{\boldmath$\nabla$} S$
    is then orthogonal to the walls, we can extend the field $\zeta $ out of the
     cavity so that it changes to its conjugate value $\bar{\zeta}$, under the reflections with respect to the walls. Indeed, as $\phi ^*\sigma =\zeta ^*\sigma$, the two scalars define the same electromagnetic field according to (\ref{10}).

The second transformation $T_2$ is
\begin{equation}
\zeta \rightarrow \phi ^{\rm new}={\zeta -i \over \zeta +i},
\label{220}
\end{equation}
that maps the half complex plane $\Im \zeta >0$ (resp. $\Im \zeta
<0$) into the interior (resp. exterior) of the unit circle $|\phi
^{\rm new}|<0$ (resp. $|\phi ^{\rm new}|>0$). Moreover, if $\zeta
_1=\bar{\zeta _2}$, then $\phi _1^{\rm new}= 1/\bar{\phi}_2^{\rm
new}$. It is easy to show that $\zeta$ and $\phi ^{\rm new}$
generate the same electromagnetic field by pull-back (\ref{10}). Indeed, it is easy to show that
$\phi ^{{\rm new}*}\sigma =\zeta ^*\sigma$ (because $d\zeta \wedge d\bar{\zeta} /(1+|\zeta |^2)^2 =d\phi ^{\rm new}\wedge d\bar{\phi}^{\rm new}/(1+|\phi ^{\rm new}|^2)^2)$, so that $\phi ^{\rm new}$ and $\zeta$ define the same electromagnetic field. We have thus showed that $\phi ^{\rm new}$ obeys the first equation (\ref{190}).

As ${\bf E}$ is orthogonal to the walls, its lines connect the cavity with its reflected images. Therefore, two points symmetric with respect to a wall which are in the same line have the same value of $\theta$. If a subset of lines form an island inside the cavity without going out of it, we can assign the same value to their images under reflections. This means that $\theta (-x_i)=\theta (x_i)$, so that the second equation (\ref{190}) has been proved also.

Note that, after applying the transformation $T_2T_1$, the resulting  Clebsch variable $s,\, u,\, v$ are even under reflection with respect to the walls of the cavity while $p-1/2$ is odd (hence $p=1/2$ in the walls). In particular $\mbox{\boldmath$\nabla$} s$ and  $\mbox{\boldmath$\nabla$} u$ are tangent to the walls, their Fourier expansions being of the form $s=\sum s_{ijk}(t) \cos ix\cos jy\cos kz$, $u=\sum u_{ijk}(t)\cos ix\cos jy \cos kz$.

We define now the scalars $\phi ^\prime$, $\theta ^\prime$ by means of equations (\ref{30a})-(\ref{30b}) applied to the two pairs of Clebsch variables $(P,\, S)$, $(V,\, U)$, respectively, given as
\begin{eqnarray}
P&=&     {1+(p({\bf r},0)-1/2)\cos \omega t\over 2}, \;\;\;\; S=2s ({\bf r} ,0),\nonumber \\
 V&=&{1+v({\bf r},\pi /2\omega )\sin \omega t\over 2}, \;\;\;\; U=2u ({\bf r} ,\pi /2\omega ).
\label{230}
\end{eqnarray}
It is clear that the symmetry properties of $P,\, S,\, U,\ V$ are the same as those of $p,\,s,\,u,\, v$ previously explained. This, together with their time dependence shows that the scalars $\phi ^\prime$ and $\theta ^\prime$ satisfy the relations (\ref{180})-(\ref{210}).

It turns out that
\begin{equation}
\mbox{\boldmath$\nabla$} P\times \mbox{\boldmath$\nabla$} S =  \mbox{\boldmath$\nabla$} (p({\bf r} ,0)\cos \omega t)\times \mbox{\boldmath$\nabla$} s ({\bf r} ,0)={\bf B}({\bf r} ,0) \cos \omega t= {\bf B}({\bf r} ,t),\label{240}
\end{equation}
\[
\mbox{\boldmath$\nabla$} U\times \mbox{\boldmath$\nabla$} V  = \mbox{\boldmath$\nabla$} (u ({\bf r} ,\pi /2\omega  )\sin \omega t)\times \mbox{\boldmath$\nabla$} v ({\bf r} ,\pi /2\omega )={\bf E}({\bf r} ,\pi /2\omega ) \sin \omega t={\bf E}({\bf r}, t). \nonumber
 \]
This shows that the new pair $\phi ^\prime ,\,\theta ^\prime$ and the old one $\phi ,\,\theta $ have something important in common: in both cases, their level curves are the magnetic and electric lines, respectively.
Let the vector fields ${\bf E}^*$, ${\bf B}^*$ be defined as
$${\bf E}^*= \partial _0 S\mbox{\boldmath$\nabla$} P -\partial _0 P \mbox{\boldmath$\nabla$} S = \omega (p({\bf r},0)-1/2)\nabla s\, \sin \omega t ,$$
$${\bf B}^* =\partial _0 U\nabla V -\partial _0 V \nabla U = -\omega v({\bf r}, \pi /2\omega  )\nabla u \cos \omega t. $$
As a consequence, the pull-backs of the area 2-form in $S^2$ by $\phi ^\prime$ and $\theta ^\prime$ are given by ${\cal F}^\prime \equiv ({\bf E}^*, {\bf B})$ and ${\cal G}^\prime \equiv (-{\bf B}^*, {\bf E})$.

Note that ${\bf E}^*$ vanishes in the walls while ${\bf B}^*$ is tangent to them, because of the properties of these Clebsch variables.
Taking the curl of the starred fields, it is seen that they obey the equations
$$\nabla \times {\bf E}^*=-{\partial {\bf B}\over \partial t},\;\;\;\; \nabla \times {\bf B}^*={\partial {\bf E}\over \partial t}.$$
This implies that the differences ${\bf B}^*-{\bf B}$ and ${\bf E}^*-{\bf E}$ have zero curl, so that
\begin{equation}
{\bf E}^*-{\bf E} =\nabla \alpha ,\;\;\;{\bf B}^*-{\bf B} =\nabla \beta ,
\label{252}
\end{equation}
where $\alpha$ and $\beta$ are space functions times $\sin \omega t$ and $\cos \omega t$, respectively. Note that $\nabla \beta$ is parallel to the walls while $\nabla \alpha $ is normal to them, $\alpha$ being therefore constant in the border of the cavity at any time.

A lemma is proved now that will be needed in section 5.

Lemma: The following equalities hold true
\begin{equation}
\int _{\cal C} \left({\bf B}\cdot {\bf B}^*\right)\, d^3r =\int _{\cal C} {\bf B}^2\, d^3r,\;\;\;\;
\int _{\cal C} \left({\bf E}\cdot {\bf E}^*\right)\, d^3r =\int _{\cal C} {\bf E}^2\, d^3r.
\label{262}
\end{equation}
The proof is simple. Integrating by parts and using the divergence theorem, the first of these integrals is equal to
$$\int _{\cal C}\left({\bf B}^2 + {\bf B}\cdot \nabla \beta \right)\,d^3r=
\int _{\cal C}{\bf B}^2\, d^3r + \int _S \beta {\bf B}\cdot {\bf n}\, da=  \int _{\cal C}{\bf B}^2\, d^3r,$$
where the fact  that ${\bf B}$ is tangent to the walls $S$ has been used. In the case of the electric field, we have
$$\int _{\cal C}\left({\bf E}^2 + {\bf E}\cdot \nabla \alpha \right)\,d^3r=
\int _{\cal C}{\bf E}^2\, d^3r + \int _S \alpha {\bf E}\cdot {\bf
n}\, da=  \int _{\cal C}{\bf E}^2\, d^3r,$$ the second integral in
the middle being zero because $\alpha$ is constant in the walls
and $\nabla \cdot {\bf E}=0$. Note that the integrands include
factors $\cos ^2\omega t$ or $\sin ^2\omega t$, respectively.

\section{Definition of two orbifolds}
 The map $\eta \equiv \phi \times \theta :T^4\rightarrow S^2\times S^2$ was considered at the beginning of section 4. The periodicity properties involved being the same, we can define now the map
\begin{equation}
\tilde{\chi}\equiv \phi ^\prime \times \theta ^\prime _{\rm Ad} :T^4\rightarrow S^2\times S^2,
 \label{262b}
\end{equation}
where $\theta ^\prime _{\rm Ad}({\bf r} ,t)=\theta ^\prime ({\bf r} , t+\tau /4)$
(note that the Clebsch variables of $\theta ^\prime _{\rm Ad}$ are $U_{\rm Ad}=U$, $V_{\rm Ad}=(1+v({\bf r},\pi /2\omega )\cos \omega t)/2$).

We will see that the map $\tilde{\chi}$ has an interesting structure. Because of the symmetry relations (\ref{180})-({\ref{210}), it turns out that $\phi ^\prime \rightarrow 1/\bar{\phi}^\prime $, under reflections with respect to the walls of the cavity $x_i=\pi$,  while $\theta ^\prime _{\rm Ad}$ remains invariant. This means that we can construct the map $\chi$ by extending the two scalars from the interior of the cavity ${\cal C}: 0\leq x,y,z\leq \pi$ to all the cube ${\cal C}^*:0\leq x,y,z\leq 2\pi$ by means of reflections, in such a way that, by applying $N$ reflections to $\phi ^\prime$, it changes to $1/\bar{\phi}^\prime$ if $N$  is odd but does not change if $N$ is even, while $\theta ^\prime _{\rm Ad}$ does not change in any case.
 On the other hand, $\theta ^\prime _{\rm Ad}$ has the same   time dependence as $\phi ^\prime $, so that it obeys the same symmetry relations with respect to the time transformations, {\em i.e.} the first equations in each of the pairs (\ref{195})-(\ref{210}).
More precisely $\theta ^\prime _{\rm Ad}$ changes into the inverse
of its complex conjugate under the changes $t\rightarrow t+\tau
/2$ and $t\rightarrow -t$ and remains invariant if $t\rightarrow
\tau /2-t$.

This suggests the convenience of identifying $\phi ^\prime $ with $1/\bar{\phi}^\prime $ and  $\theta ^\prime _{\rm Ad}$ with $1/\bar{\theta}^\prime _{\rm Ad}$ and of defining two equivalence relations, ${\cal R}_{\rm f}$ and ${\cal R}_{\rm st}$, between pairs of fields and between spacetime points as follows.

a) The two pairs $(\phi ^\prime ,\theta ^\prime _{\rm Ad})$ and  $(\phi ^\dag ,\theta ^\dag _{\rm Ad} )$ are equivalent according to ${\cal R}_{\rm f}$, if either $\phi ^\prime =\phi ^\dag$ or $\phi ^\prime =1/\bar{\phi} ^\dag$ and  either $\theta ^\prime _{\rm Ad}=\theta _{\rm Ad}^\dag$ or $\theta ^\prime _{\rm Ad}=1/\bar{\theta} _{\rm Ad}^\dag$.

b) The two spacetime points $(x_k, t)$ and $(x_k^\prime , t^\prime)$ in $T^4=T^3\times T^1$ are equivalent according to ${\cal R}_{\rm st}$ if one can go from the one to the other by means of one or several reflections with respects to the planes $x_k=\pi$ and $t=\tau /4$ or $t=\tau /2$.

These two equivalence relations are interesting because of the symmetry relations (\ref{180})-(\ref{210}) as will be seen now.
Let us consider two groups $G_{\rm t}$, $G_{\rm s}$ acting on the time and space variables in $T^4$, respectively, defined as follows. The group $G_{\rm t}$, of order 4, is generated by the two  symmetries with respect to the times $t=\tau
/4$ and $t=0$, {\em i.e.} ${\cal T}_1: t\rightarrow
\tau /2 -t$  and ${\cal T}_2: t\rightarrow -t \equiv \tau -t$, respectively.  The product ${\cal L}= {\cal T}_2{\cal T}_1$ is the
translation of length $\tau /2$. If $\cal I$ is the identity, then $G_{\rm t}\equiv \{{\cal I},\; {\cal T}_1,\; {\cal T}_2,\,
{\cal L}\}$. It is a representation of the dihedral group $D_2$. The images of any time by the elements of this group are four times, each one in a different quarter of period.

The group $G_{\rm s}$ is the set of the 8 space transformations generated by the reflections with respect to the walls of the cavity $G_{\rm s}\equiv\{ {\cal I}, {\cal S}_1, {\cal S}_2,{\cal S}_3,{\cal L}_1,{\cal L}_2,{\cal L}_3,{\cal S}_{123}\}$, where $\cal I$ is the identity; ${\cal S}_i$, the reflection with respect to the wall $x_i=\pi$; ${\cal L}_i=\epsilon _{ijk}{\cal S}_j{\cal S}_k$, the rotation of $\pi$ around the axis $x_j=x_k=\pi$ and ${\cal S}_{123}={\cal S}_1{\cal S}_2{\cal S}_3$, the symmetry with respect to the point $(\pi ,\pi ,\pi)$. The images of any space point of $T^4$ by the 8 elements of this group are 8 points, one in the cavity and the other seven in each one of the seven cubes of side $\pi$ obtained from the cavity by symmetries with respects to the walls.

Considering now both space and time, the torus $T^4$ is the union
of 32 subsets, each one being the cartesian product of one quarter
period of time $\times$ one of the eight cubes of side $\pi$. The
symmetry properties involve the group
$G=G_{\rm s}\times G_{\rm t}$, of 32 elements ($=8\times 4$), its action in $T^4$ being $g
({\bf r},t)= (g_{\rm s}{\bf r}, g_{\rm t}t)$, where $g=g_{\rm s}\times g_{\rm t} \in G,\;
g_{\rm s}\in G_{\rm s},\; g_{\rm t} \in G_{\rm t}$).
Its 32 elements transform any subset in
itself plus the other 31 subsets. The important point to
stress is that the equivalence classes of
${\cal R}_ {\rm st}$ are the orbits of the spacetime points by the action of $G$, {\em i.e.} the 32 spacetime points obtained from any
point. The corresponding values of the two scalars are in the
same equivalence class of ${\cal R}_{\rm f}$. In other words, we
have a spacetime orbifold, a field orbifold and a map between
them.

These two orbifolds are defined as the quotients of the torus $T^4$ by ${\cal R}_{\rm st}$ and of the product $S^2\times S^2$ by ${\cal R}_{\rm f}$, respectively. Noting them $O^4$ and $\Sigma ^4$, we have
 \begin{equation}
O^4\equiv T^4 /{\cal R}_{\rm st},\;\;\;\; \;\;\;\; \Sigma ^4 \equiv \left( S^2\times S^2\right) /{\cal R}_{\rm f}.
\label{270}
\end{equation}
It is clear that  $\Sigma ^4=D^2 \times D^2$, where $D^2$ is the 2-disk.

 This means that, in the TME, a normal mode of the electromagnetic field in a cubic cavity must be defined by a map $\chi$ from $O^4$ to $\Sigma ^4$ of the form
\begin{equation}
\chi =\phi ^\prime \times \theta ^\prime _{\rm Ad}: O^4\rightarrow \Sigma ^4.
\label{280}
\end{equation}
Conversely, any such map defines a periodic electromagnetic field in the cavity with frequency $\omega$, if the complex scalar fields $\phi (t)$ and $\theta (t)=\theta _{\rm Ad}(t-\tau /4)$, from which $\phi ^\prime$ and $\theta _{\rm Ad}^\prime$ have been constructed, are dual in the sense of the condition (\ref{20}). It must be emphasized that  this duality condition is necessary for the validity of the main result of this work.

As the previous considerations show, we can take any one of the 32 subsets of $T^4$ as the basic manifold of $O^4$, with due attention to the orientation, for instance the cavity times a quarter period, say  ${\cal C}\times [0,\tau /4]$. Indeed, $O^4$ can be thought of as the 32 subsets folded one on top of each other, half of them with the opposite orientation as the other half. As basic manifold of  $\Sigma ^4$, we can take  the product of two unit disks $D^2\times D^2$, defined by the conditions $|\phi ^\prime |\leq 1$ and $|\theta _{\rm Ad}^\prime|\leq 1$.

\section{Topological quantization of the energy}
We can represent the maps and projections on the quotients by the
diagram \newpage \begin{equation} \tilde{\chi}:T^4 \;\;\;
\stackrel{\tilde{d}}{\rightarrow} \;\;\; S^2\times S^2
\label{280a}
\end{equation}
$$\!\!\!\!\!\!\!\! \downarrow \;\;\;\;\;\;\;\;\;\;\;\;\; \;\;\;\;\downarrow $$
\begin{equation}
\chi:O^4\equiv T^4/{\cal R}_{\rm st},\;\;\;\;
\stackrel{d}{\rightarrow} \;\;\;\; \Sigma ^4 \equiv S^2\times
S^2/{\cal R}_{\rm f}, \label{280b}
\end{equation}
where $\tilde{d}$ and $d$ are two integers equal to the degrees of the corresponding maps. It turns out therefore that \cite{Mil65,Gui74}
\begin{equation}
d= \int _{0}^{\tau /4} \int _{\cal C} 4\phi ^{\prime \,*}\sigma \wedge \theta _{\rm Ad}^{\prime \,*}\sigma
=\int _{O^4} 4{\cal F}^\prime \wedge {\cal G}^\prime _{\rm Ad},
\label{290}
\end{equation}
where the integrand is the normalized area 2-form in $D^2\times D^2$. Note (i) the limits of the time integral (because it must be extended to the basic manifold of $O^4$ which is ${\cal C}\times [0,\tau /4]$, as discussed at the end of the previous section); and (ii) the factor 4 in the normalized area 2-form $4\phi ^{\prime \,*}\sigma \wedge \theta _{\rm Ad}^{\prime \,*}\sigma$ (since $D^2\times D^2\, (\equiv S^2\times S^2/{\cal R}_{\rm f})$ is a fourth part of $S^2\times S^2$). This implies that \begin{equation}
d=4\int _{0}^{\tau /4}dt \int _{\cal C} {E(t)^*E(t+\tau /4)-B(t)B^*(t+\tau /4)\over 2}\, d^3x.
\label{300}
\end{equation}

Note an important point\footnote{I am indebted to referee 1 for
suggesting me this aspect of the question.}: the integral
(\ref{170}) vanishes because it extends over the 32 subsets of
$T^4$, the orientation of 16 of them being the opposite to that of
the other 16. Indeed, if the integrals (\ref{290})-(\ref{300})
were extended over one period instead of over one quarter, they
would vanish, since the four integrals over the quarters have
equal moduli, two being positive and  two negative. However, a map
with nonvanishing topological index is defined by means of the
orbifold construction. Indeed, one could write instead of
(\ref{300})
\begin{equation}
d=\int _{0}^\tau  \left| \int _{\cal C} {E(t)^*E(t+\tau
/4)-B(t)B^*(t+\tau /4)\over 2}\, d^3x \right|dt. \label{305}
\end{equation}
This shows that, when the orbifold construction is used, the
information of the orientation of the map $\tilde{\chi}$ is lost,
this being curiously what gives information on the modulus of
(\ref{300}): the integral no longer vanishes but is an integer
number. Indeed the map between the orbifolds can be understood as
the restriction of $\tilde{\chi}$ to the basic cell ${\cal
C}\times [0,\tau /4]$, since the maps (\ref{280a}) and
(\ref{280b}) only describe the same map in that basic cell, so
that the loss of information when passing from the first to the
second is essential for the proof of (\ref{300}).

Because of the lemma of the previous section we can substitute in (\ref{290})
${\bf E}^*$ for ${\bf E}$ and ${\bf B}^*$ for ${\bf B}$ in the
last integral  (the shift in time does not affect this
result). From equations (\ref{70})-(\ref{80}), it turns out that
$$ \int _{\cal C} {E(t)E(t+\tau /4)-B(t)B(t+\tau /4)\over 2}\, d^3x =\omega ^2A^2{\pi ^3\over 16}\sin2\omega t={\cal E}\sin 2\omega t,$$
 where $\cal E$ is the energy, given by equation (\ref{120}), so that
$$d=4\int _{0}^{\tau /4}{\cal E}\sin 2\omega t\, dt =4{\cal E}\left.{-\cos 2\omega t\over 2\omega}\right|_{0}^{\tau /4},$$
which can be written as
\begin{equation}
{\cal E}=n\omega ,
\label{310}
\end{equation}
with $n=d/4$ (remember: $d$ is an integer) or, equivalently,
${\cal E}=\int _{\cal C}(E^2+B^2)d^3r/2= (d/4)\omega$ (or ${\cal
E}=(d/4)\hbar \omega$ in physical units). The main result of this
work is equation (\ref{310}), which establishes a topological
quantization of the energy. The number of photons $n$ is here one
fourth of the degree $d$ of a map between two orbifolds. This
means that the energy is quantized and is always an integer
multiple of $\omega /4$ (of $\hbar \omega/4$ in physical units).

 We see that the topological model  proposed in \cite{Ran90a,Ran92,Ran95b}
implies and embodies a topological quantization law of the energy
in a cubic cavity, similar to that of Planck and Einstein but with
an extra factor $1/4$. An important question is whether there is
any reason for $d$ to be be a multiple of four. In that case
equation (\ref{310}) would coincide not only qualitatively but
also quantitatively with Planck-Einstein law, and the TME would
predict the correct law for the quantization of the energy of the
radiation in a cubic cavity. This question will be considered in a
future paper.

\section{Final comments and conclusion}
Since Lagrange's time, physicists have the freedom of using different sets of coordinates to describe any system in equivalent ways. The topological model considered here is based on the idea of magnetic and electric force line \cite{Ran90a,Ran92,Ran01}, its natural coordinates being pairs of complex scalar fields, the level curves of which being these force lines.  The only additional assumption that must be made to develop the model is the compactification of the space $R^3$ to $S^3$, which allows to interpret these scalars as maps $S^3\rightarrow S^2$ (in open 3-space) or $T^3\rightarrow S^2$ (in a cubic cavity), with the immediate consequence that the Faraday 2-form and its dual become the pull-backs of the area 2-form in $S^2$, this being the main reason for the rich topological properties. After that all goes straightly (if not easily). Although the TME is locally equivalent to Maxwell standard theory, it embodies besides the topological quantization of the electromagnetic helicity (\ref{40}), with the meaning of a Hopf index \cite{Ran95b,Ran01} expressing the linking of the force lines, and the topological quantization of the electric charge. This paper shows that the TME embodies also the topological quantization in a cubic cavity.  As the model makes use of only  {\em c}-number fields in its present  still non quantized form, it may seem surprising to find that it embodies a relation usually  considered to be the very basis of quantum theory. In spite of that, the present paper  must not be interpreted in any way whatsoever as an attempt to reduce quantum physics to classical terms. It is not.

On the contrary, the evidence that quantum mechanics is here to stay is not disputed here, that would be an impossible task certainly doomed to failure. What equation (\ref{310}) shows is a different thing: that the discretization of a physical quantity, as the energy or any other, might be of a different nature from other quantum properties as the interference of amplitudes, which cause the probabilities not to follow the Laplacian rules, the entanglement, or the zero point energy\footnote{This was already pointed out in other contexts, regarding the discrete ``quantum" numbers  arising in solitons or in the quantization of the electric charge and gluon charge because of their identification with the Chern class of a fiber bundle.}. These last three properties represent much more radical departures from classical ideas. In fact, there are many quantities in classical physics that take only discrete values, including for instance the pitch of strings, drums or tubes of musical instruments. It is ironic that the very name ``quantum physics"  comes from the less radical novelty introduced by the theory. There is some ground to argue that this has added a bit of confusion to some debates on the foundations of quantum mechanics. Maybe that name was not the most adequate election.

In conclusion,  this paper gives support to the topological model of electromagnetism with its electromagnetic knots and to the mechanisms of charge quantization and helicity quantization that it implies. It must be further investigated since it could give some further insight into the interplay of the quantization process of the electromagnetic field and its topological properties.

\section{Acknowledgments}
I am indebted to Dr. J. L. Trueba and Prof. A. Tiemblo for
discussions and encouragements and to Profs. J. M. Montesinos and
A. Ibort for help on the mathematical aspects of this work.  I am
also thankful to referee 1 for his/her helpful criticism.

\end{document}